\documentclass[sigconf,anonymous=False]{acmart}

\usepackage{booktabs} 
\usepackage[english]{babel}
\usepackage{subcaption}
\usepackage{tabularx}
\usepackage[inline]{enumitem}
\usepackage{multirow}

\newenvironment{ppl}{\fontfamily{ppl}\selectfont}{}

\begin{document}

\title[]{WikiPassageQA: A Benchmark Collection for Research on Non-factoid Answer Passage Retrieval}

\author{Daniel Cohen \quad Liu Yang \quad W. Bruce Croft}
\affiliation{%
	\institution{
	Center for Intelligent Information Retrieval, University of Massachusetts Amherst, Amherst, MA, USA}
}
\email{{dcohen, lyang, croft}@cs.umass.edu}

\renewcommand{\shortauthors}{D. Cohen et al.}

\begin{abstract}
\noindent With the rise in mobile and voice search, answer passage retrieval acts as a critical component of an effective information retrieval system for open domain question answering. Currently, there are no comparable collections that address non-factoid question answering within larger documents while simultaneously providing enough examples sufficient to train a deep neural network. In this paper, we introduce a new Wikipedia based collection specific for non-factoid answer passage retrieval containing thousands of questions with annotated answers and show benchmark results on a variety of state of the art neural architectures and retrieval models. The experimental results demonstrate the unique challenges presented by answer passage retrieval within topically relevant documents for future research. 
\end{abstract}

\copyrightyear{2018} 
\acmYear{2018} 
\setcopyright{acmcopyright}
\acmConference[SIGIR'18]{The 41st International ACM SIGIR Conference on Research & Development in Information Retrieval}{July 8--12, 2018}{Ann Arbor, MI, USA}
\acmBooktitle{SIGIR'18: The 41st International ACM SIGIR Conference on Research & Development in Information Retrieval, July 8--12, 2018, Ann Arbor, MI, USA}
\acmPrice{15.00}
\acmDOI{10.1145/3209978.3210118}
\acmISBN{978-1-4503-5657-2/18/07}

\fancyhead{}
\settopmatter{printacmref=false, printfolios=false}


\maketitle


\section{Introduction}

Recent advances in deep learning have allowed recent work in numerous fields to achieve state of the art performance on key tasks, with larger networks often outperforming smaller networks after accounting for overfitting. However, these deep neural networks contain millions of parameters even with only a small number of layers that necessitates a large amount of training data compared to more conventional models. As such, high quality openly available benchmark data sets are critical for research progress.   Examples include ImageNet \cite{Deng09imagenet} for computer vision and SQuAD \cite{DBLP:journals/corr/RajpurkarZLL16} for machine comprehension. Large, high quality datasets allow the community to not only rapidly develop new models for a task, but also to iteratively learn how a model architecture learn better representations for a specific task.

With the rising popularity of mobile and voice assisted search, where the size of screen and the output length is limited, there is a growing need to develop models for retrieving answer passages. Here, the information need of a query lies between that of a short fact or single sentence, and a document, and cannot be sufficiently answered with either. In terms of question answering, there are existing datasets such as TREC QA \cite{wang-smith-mitamura:2007:EMNLP-CoNLL2007}, WikiQA \cite{yang-yih-meek:2015:EMNLP} and InsuranceQA \cite{DBLP:journals/corr/FengXGWZ15} that provide sufficient collections of queries to train a neural network \cite{DBLP:journals/corr/YuHBP14,DBLP:conf/acl/WangN15,Severyn:2015:LRS:2766462.2767738,Yang:2016:ARS:2983323.2983818}.
However, these datasets do not address the answer passage retrieval task since their focus is on retrieving factoids, short snippets, or isolated sentences.
In the collection introduced by this paper, the task is not only to retrieve a passage that answers the question, but also to identify where the answer portion of a document begins and ends within a larger topically relevant document.

Currently, there is only one collection specifically created for retrieving answer passages in documents, WebAP \cite{Keikha:2014:EAP:2600428.2609485}, where contiguous sentences of a document are labeled as relevant to a query. While addressing the answer passage retrieval task, the WebAP collection suffers from a small number of queries, resulting in poor performance of neural models.

In this paper we present a new collection, WikiPassageQA, containing $4,165$ queries created from Amazon mechanical turk\footnote{\url{https://www.mturk.com/mturk/welcome}} over the top $863$ Wikipedia documents from the Open Wikipedia Ranking\footnote{http://law.di.unimi.it/}. Each Wikipedia page has multiple queries accompanied with locations of varying length answer passages within the document. 

The contributions of this work are as follows: (1) We introduce a new benchmark collection for the research on non-factoid answer passage retrieval\footnote{It can be downloaded from \url{https://ciir.cs.umass.edu/downloads/wikipassageqa}}. 
(2) We perform extensive experiments with WikiPassageQA to show benchmark results of various methods including traditional and neural IR models that demonstrate the unique challenges that differentiate answer passage retrieval from past QA tasks. 

\section{Existing Related Datasets}

We perform a survey of related question answering and reading comprehension data sets to highlight the differences between them and WikiPassageQA.

\textbf{Factoid Question Answering. }There are several benchmark data sets for the evaluation of factoid question answering, which aim to identity short answer facts such as named entities, numbers and noun phrases. Wang et al. \cite{wang-smith-mitamura:2007:EMNLP-CoNLL2007} developed a benchmark collection using the Text REtrieval Conference (TREC) 8-13 QA data. They used the questions in TREC 8-12 for training and set aside TREC 13 questions for development (84 questions) and testing (100 questions). This TREC QA data set has become one of the most widely used benchmarks for answer sentence selection \cite{DBLP:journals/corr/YuHBP14,DBLP:conf/acl/WangN15,Severyn:2015:LRS:2766462.2767738,Yang:2016:ARS:2983323.2983818}. Recently, Yang et al. \cite{yang-yih-meek:2015:EMNLP} created the WikiQA dataset using Bing query logs and Wikipedia passages as the source of answers. WikiQA data is more than an order of magnitude larger than the previous TREC QA data. Feng et al. \cite{DBLP:journals/corr/FengXGWZ15} created InsuranceQA, which is a data set in the insurance domain. It consists of questions from real world users and answers composed by professionals with deep domain knowledge about insurance. These data sets either only include very short answers and answer sentences for factoid questions, or only for a closed domain like insurance. However, the WikiPassageQA data proposed in this paper includes many long passages for non-factoid questions and there are no restricted domains for these questions and answers.

\textbf{Non-Factoid Question Answering.} There have been previous efforts on developing benchmark data sets for non-factoid question answering or answer passage retrieval~\cite{Keikha:2014:EAP:2600428.2609485, Habernal:2016:NCA:2911451.2914682, DBLP:conf/ecir/YangASCPCGS16}. Perhaps the closest prior research to our work is the WebAP data set created by Keikha et al.~\cite{Keikha:2014:EAP:2600428.2609485,DBLP:conf/ecir/YangASCPCGS16}. Compared to WebAP, WikiPassageQA has a two significant differences: (1) the number of questions in WikiPassageQA is significantly larger than that of WebAP (4165 v.s. 82). 
(2) WikiPassageQA has different properties on the specificity of queries. WebAP used previous TREC tropical queries whereas WikiPassageQA has questions with more focused information needs.


There are also non-factoid QA data built from community question answering (CQA) data. The most commonly known of these are the Yahoo L4 ``manner'' questions and a filtered non-factoid collection from the entire Yahoo L6 Webscope collection (nfL6)\cite{cohen-ictir}. While both CQA collections and WikiPassageQA target non-factoid questions, there are two significant differences between them. 

(1) The candidate answers from the CQA collections either come from other questions, which may not have any semantic relationship to the target query, or come from ``non-best'' answers submitted in response to the query. These candidate answers have unreliable and generally missing labels. (2) As opposed to WikiPassageQA, these CQA collections consist of answer passages without surrounding text. These two factors separate CQA collections from that of the task defined by the collection proposed in this paper. 


\textbf{Reading Comprehension.} The other related data sets are reading comprehension data sets including MCTest \cite{richardson-burges-renshaw:2013:EMNLP}, CNN /Daily News \cite{DBLP:journals/corr/HermannKGEKSB15}, Children's Book Test \cite{DBLP:journals/corr/HillBCW15}, SQuAD \cite{DBLP:journals/corr/RajpurkarZLL16}, MS MARCO \cite{DBLP:journals/corr/NguyenRSGTMD16}, BAbI \cite{DBLP:journals/corr/WestonBCM15}, etc. Unlike answer sentences or passages in the question answering datasets, these reading comprehension data sets mostly involve selecting a specific short span within a sentence, selecting an answer from predefined choices, or predicting a blanked-out word of a sentence given previous context sentences. WikiPassageQA stands apart by using only user annotated answer passages rather than synthetic data, and most accurately reflects the task of finding raw answer passages within a larger document.
 

In summary, WikiPassageQA is the only large data set with long passages as answers for thousands of non-factoid questions in the open domain.

\section{The WikiPassageQA Dataset}

 \begin{table}[htbp]
 	\caption{WikiPassageQA collection statistics. ``P'' in the first column denotes ``Passages''.} 
\centering
\begin{tabular}{l|l|l|l|l}
\hline \hline
Data & Train & Dev & Test & Total\\ 
\hline
Questions & 3332 & 417 & 416 & 4165\\ 
\hline
CandidateP & 194314 & 25841 & 23981 & 244136\\ 
\hline
PosCandidateP  & 5560 & 707 & 700 & 6967\\ 
\hline
NegCandidateP  & 188754 & 25134 & 23281 & 237169\\ 
\hline
\% of PositiveP & 0.049 & 0.043 & 0.051 & 0.049\\ 
\hline
CandidateP/Query & 58.318 & 62.968 & 57.647 & 58.616\\ 
\hline
PosCandidateP/Query & 1.669 & 1.695 & 1.683 & 1.672\\ 
\hline
AvgLenOfQuestion & 9.52 & 9.69 & 9.44 & 9.53\\ 
\hline
AvgLenOfAnswerP & 133.092 & 134.132 & 132.650 & 133.158\\ 
\hline
\hline\end{tabular}
\label{table:stats}
\end{table}

\subsection{Query And Answer Passage Synthesis}

The dataset was created using Amazon's mechanical turk platform, where we sourced high quality crowd workers to create questions based on a Wikipedia document. We restricted workers to have over 1000 assignments completed as well as having over a 98\% approval rating to ensure quality submissions. While workers were able to work on multiple human intelligence tasks (HITs), no worker was able to submit twice on the same Wikipedia page.
In a similar manner to the creation of the SQuAD collection~\cite{DBLP:journals/corr/RajpurkarZLL16}, each worker was asked to create five non-factoid questions and indicate location of their respective answer passages within the document. ``Who'', ``Where'', and ``When'' questions were explicitly prohibited to prevent factoid answers. A relevant passage was deemed to be more than one contiguous sentence, with no additional information that doesn't address the query. In order to prevent low quality submissions, workers were able to submit less than five queries if the document was not suitable for the task. Workers were paid \$0.65 per HIT and the total cost of data annotation was \$638.

\subsection{Evaluating Answer Passage Quality}
Once a batch of question and answer passages was completed, they were resubmitted to the Amazon mechanical turk platform in a verification poll. For each question and assignment passage from an assignment, five workers were asked to provide two ratings: (1) rate the question as \textit{factoid}:$~0$, \textit{non-factoid}:$~1$ and  (2) the answer passage as \textit{Excellent, Great, Fair, Poor} with point values $3,2,1,0$ respectively. The Kappa coefficient of question type was 0.930 and 0.659 for factoid/non-factoid and answer passage quality during this evaluation process, which indicates good agreement score among different annotators.
Question-answer passage pairs were removed if mean scores for these two ratings were, respectively, less than 0.66 and 2 to ensure quality. This filtering process reduced the original collection of question-answer passage pairs from 4908 to 4165 pairs. 

\subsection{Collection Characteristics}

\begin{figure}[h]
	\caption{Distribution of Answer Passage Lengths.}
	\centering
	\includegraphics[width=0.45\textwidth]{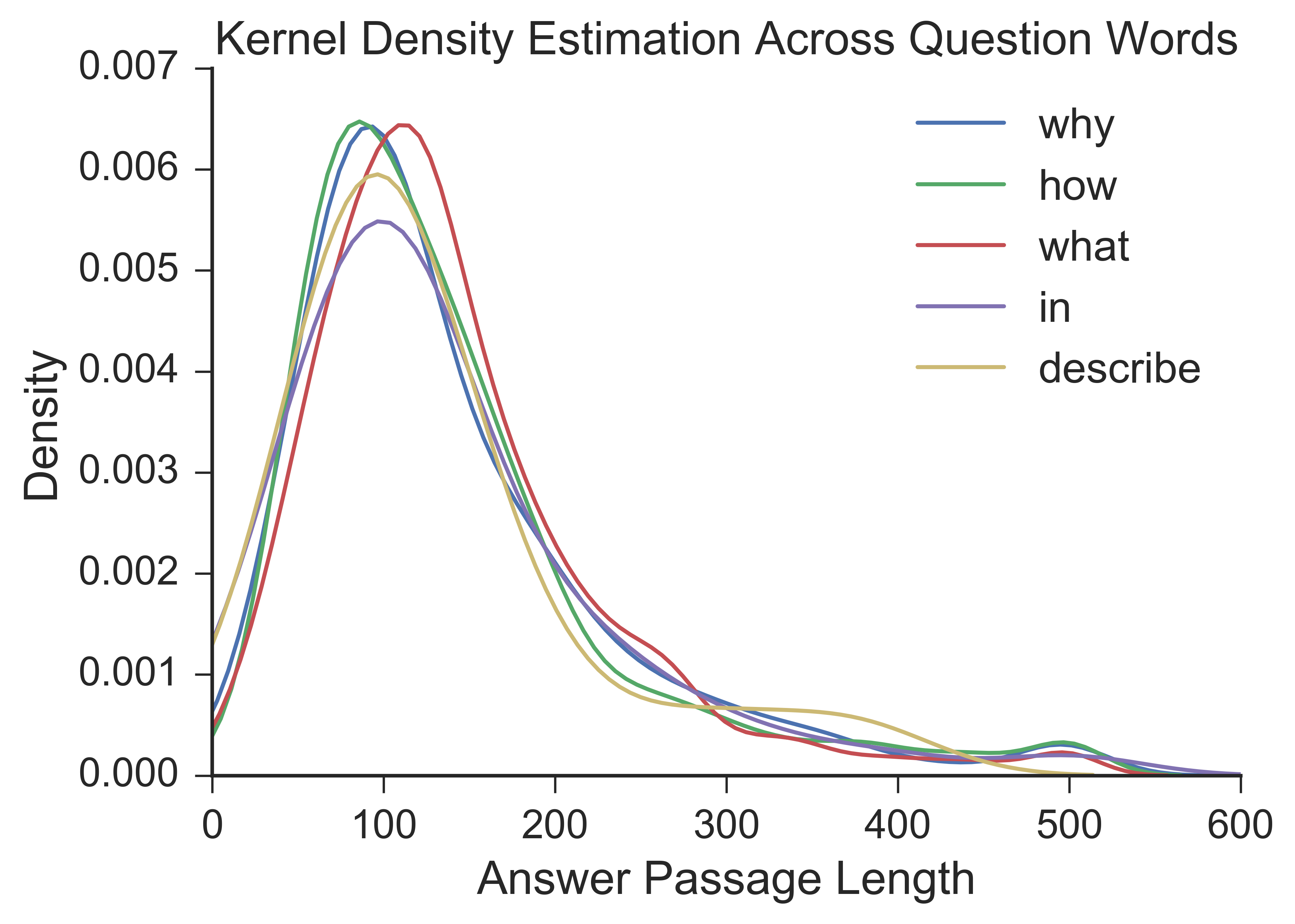}
	\label{fig:data}
\end{figure}

As seen in Table~\ref{table:stats}, the filtered collection possesses annotated answer passages significantly longer than previous QA datasets. Breaking down the queries by the first word of the question, ``what'', ``how'' and ``why'' make up 43.8\%, 36.6\%, and 14.0\% of the collection. The next most common start word is ``in'' at 1.2\%, acting as a prepositional phrase for the question. Across all question words, Figure~\ref{fig:data} shows that the answer passages have a similar length distribution with 99.9\% of all passages having less than 400 words. As there is only one relevant passage for each question, there is a risk of false negative passages. However, due to the specific prompt of requiring the information need of the query spread over multiple sentences, the relevant passages are highly likely to be unique to the Wikipedia page.
A comparison of sample question and annotated answers is provided in Table \ref{table:sampleQAs} between WikiPassageQA and TREC QA data.


\begin{table}[htbp]
	\centering
	\caption{Comparison of example questions and answers in TREC QA Track data and WikiPassageQA data.}
	\label{table:sampleQAs}
	\scriptsize
	\begin{ppl}
		\begin{tabular}{|p{8cm}|} 
			\hline
			\multicolumn{1}{|c|}{Sample Questions and Answers in TREC QA Track Data}\\
			\hline
			\underline{{\emph{Query 201}}}:\newline 
			Question: What was the name of the first Russian astronaut to do a spacewalk? \newline 
			Answer: Aleksei A. Leonov \newline
			Answer Document ID: LA072490-0034 \newline
			\underline{{\emph{Query 202}}}:\newline 
			Question: Where is Belize located? \newline 
			Answer: Central America \newline
			Answer Document ID: FT934-14974 \newline
			\\
			\hline
			\multicolumn{1}{|c|}{Sample Questions and Answer Passages in WikiPassageQA Data}\\
			\hline
			\underline{{\emph{Query 4114}}}:\newline 
			Question: Why is Japan so densely populated? \newline 
			Document ID: 496\\
			Document Name: Japan.html\\
			Answer Passages: \\
			The main islands, from north to south, are Hokkaido, Honshu, Shikoku and Kyushu. The Ryukyu Islands, which include Okinawa, are a chain to the south of Kyushu. Together they are often known as the Japanese archipelago. About 73\% of Japan is forested, mountainous, and unsuitable for agricultural, industrial, or residential use. As a result, the habitable zones, mainly located in coastal areas, have extremely high population densities. Japan is one of the most densely populated countries in the world.
			\newline
			\underline{{\emph{Query 2402}}}:\newline 
			Question: What is the structure of Australia's members of parliament? \newline 
			Document ID: 400\\
			Document Name: Member\_of\_parliament.html\\
			Answer Passages: \\
			\textbf{Passage 1} A Member of Parliament is the representative of the voters to a parliament. In many countries with bicameral parliaments, this category includes specifically members of the lower house, as upper houses often have a different title. Members of parliament tend to form parliamentary groups with members of the same political party. The Westminster system is a democratic parliamentary system of government modelled after the politics of the United Kingdom. This term comes from the Palace of Westminster, the seat of the Parliament of the United Kingdom. A member of parliament is a member of the House of Representatives, the lower house of the Commonwealth parliament. Members may use ``MP'' after their names; ``MHR'' is not used, although it was used as a post-nominal in the past.
			
			\textbf{Passage 2} A member of the upper house of the Commonwealth parliament, the Senate, is known as a ``Senator''. In the Australian states of New South Wales, Victoria and South Australia, a Member of the Legislative Assembly or ``lower house,'' may also use the post-nominal ``MP.'' Members of the Legislative Council use the post-nominal ``MLC.'' Members of the Jatiyo Sangshad, or National Assembly, are elected every five years and are referred to in English as members of Parliament. The assembly has directly elected 300 seats, and further 50 reserved selected seats for women. The Parliament of Canada consists of the monarch, the Senate , and the House of Commons.\\
			\hline
		\end{tabular}
	\end{ppl}
	\vspace{-0.0cm}
\end{table}

\vspace{-0.2cm}

\begin{table*}[]
\centering
\caption{Benchmark results of different methods on WikiPassageQA. Numbers in bold font mean the result is better compared with the best baseline. } 
\label{table:results}
\begin{tabular}{l|l|l|l|l|l|l|l|l|l}
\hline \hline
Type                            & Method            & MAP    & MRR    & P@5    & P@10   & nDCG   & Recall@5 & Recall@10 & Recall@20 \\ \hline \hline
\multirow{2}{*}{Base}       & WC         & 0.3456 & 0.4004 & 0.1370 & 0.0923 & 0.5096 & 0.4618   & 0.6079    & 0.7615    \\ \cline{2-10} 
                                & WC.IDF & 0.3417 & 0.3898 & 0.1351 & 0.0928 & 0.5049 & 0.4518   & 0.6129    & 0.7526    \\ \hline \hline
\multirow{3}{*}{Traditional IR} & VSM            & 0.3970 & 0.4588 & 0.1476 & 0.0921 & 0.5490 & 0.4837   & 0.5979    & 0.7464    \\ \cline{2-10} 
                                & BM25              & 0.5373 & 0.6258 & 0.1947 & 0.1151 & 0.6659 & 0.6334   & 0.7311    & 0.8309    \\ \cline{2-10} 
                                & QL                & 0.5436 & 0.6338 & 0.1947 & 0.1151 & 0.6715 & 0.6353   & 0.7275    & 0.8426    \\ \hline \hline
\multirow{5}{*}{Neural IR}      & LSTM              & 0.3352 & 0.3947 & 0.1197 & 0.0780 & 0.4912 & 0.3915   & 0.5894    & 0.7169    \\ \cline{2-10} 
                                & CNN+TF            & 0.4009 & 0.4581 & 0.1572 & 0.1099 & 0.5577 & 0.5212   & 0.7024    & 0.8412    \\ \cline{2-10} 
                                & LSTM-CNN+TF       & 0.3577 & 0.4156 & 0.1351 & 0.0942 & 0.5196 & 0.4538   & 0.6187    & 0.7608    \\ \cline{2-10} 
                                & Char+WordCNN-LSTM  & 0.4385 & 0.5534 & 0.1728 & 0.1104 & 0.5837 & 0.5709   & 0.6931    & 0.8326    \\ \cline{2-10} 
                                & Memory-CNN-LSTM+TF   & \textbf{0.5608} & \textbf{0.6792} & \textbf{0.2083} & \textbf{0.1228} & \textbf{0.6791} & \textbf{0.6522}   & \textbf{0.7329}    & \textbf{0.8592}    \\ \hline \hline
\end{tabular}
\end{table*}

\subsection{Data Overview and Experimental Settings}

As the collection consists of queries and relevant contiguous sentences, there are a variety of ways to evaluate models. In order to benchmark the dataset on common IR models for answer passage retrieval, we segment each Wikipedia article into passages of six sentences each, which is the average number of sentences in all annotated answer passages. As relevant passages can be split between windows, we deem a candidate passage as relevant if greater than 15\% of the bigrams within the annotated answer passage occurs in a candidate passage. This results in an average of 1.66 relevant passages for each query. As each Wikipedia document is distinct, at retrieval time only candidate passages from the target query's Wikipedia page were used in training and evaluation rather than all passages in the entire collection. Training of the neural models were done with a 0.8/0.1/0.1 split for training, development, and testing sets resulting in 3332, 417, and 416 queries for each set.
As this is an IR task, we partition the queries rather than Wikipedia articles common in reading comprehension tasks~\cite{DBLP:journals/corr/RajpurkarZLL16}.

\subsection{Learning Models and Evaluation Metrics}
We benchmark our dataset on two naive baselines, three traditional IR methods, and five deep neural models as shown in Table~\ref{table:results}.\\
\indent\textbf{Baselines.} These two methods (WC, WC.IDF) examine the performance using overlapped word count statistics between the question and the candidate answer passage to provide a reference point for other methods. WC.IDF is the overlapped word count statistics weighted by IDF. It can be viewed as an unnormalized TF-IDF summation.\\
\indent\textbf{Traditional IR Models.} The traditional IR models include the TF-IDF Vector Space Model (VSM), BM25 \cite{Robertson:1994:SEA:188490.188561}, and Query Likelihood (QL) \cite{Ponte:1998:LMA:290941.291008} with Dirichlet smoothing. These models will show the performances of traditional IR baselines for answer passage retrieval.\\
\indent\textbf{Neural IR Models.} Five neural models are used to evaluate answer passage retrieval with this collection: \textbf{(1)} A standard two layer \textbf{LSTM} network~\cite{cohen-ictir,DBLP:conf/acl/WangN15} is used as a simple model to benchmark a strong non factoid neural model. \textbf{(2)} \textbf{CNN+TF} adopts siamese convolutional neural networks to learn representations of questions and candidate answer passages. The QA pairs are concatenated along with \textit{tf} information after the CNN subnetwork, and passed through a feedforward network to produce a scalar relevance score, which is the approach proposed by Severyn and Moschitti \cite{Severyn:2015:LRS:2766462.2767738}. \textbf{(3) LSTM-CNN+TF} adds a LSTM layer for the long term dependency modeling prior to a CNN~\cite{tan}. This approach reflects the impact of explicitly modeling the passage as a temporal structure on ranking. \textbf{(4) Char+Word-CNN-LSTM} possesses the same structure as (2), but utilizes character embeddings to deal with out of vocabulary instances. \textbf{(5)~Memory-CNN-LSTM-TF} model uses a doc2vec~\cite{d2v} representation as its starting memory tensor, and iteratively reads and writes from it at each sentence within a passage. This includes \textit{tf} information at the sentence level, and takes into account the probability of the sentence generating each term as well.
 
 
 
 

\subsection{Experimental Results and Analysis}
As seen in Table \ref{table:results}, traditional IR models like QL achieve a very competitive baseline, outperforming all but one of the neural models. Memory-CNN-LSTM-TF outperform all other methods including traditional IR models and neural IR models. Only Memory-CNN-LSTM-TF was developed for answer passage retrieval of this length, where it sequentially iterates through each sentence while updating a memory tensor. All other neural models were designed for retrieving either sentences or passages with a mean approximate length of 50 tokens. This contrasts sharply with the characteristics of WikiPassageQA, shown in Table ~\ref{table:stats}, where the mean length of an answer passage is 142.7 tokens. Similar to the results shown in ~\cite{cohen-ictir}, CNN+TF fails to outperform a standard BM25 baseline, indicating the difficulty of neural IR architectures generalizing to new tasks at a different text granularity. The relatively poor performance of these conventional neural networks indicates the unique challenges present in the non-factoid answer passage retrieval task. The WikiPassageQA collection provides an open benchmark data with answer correctness judgments to the research community for non-factoid answer passage retrieval. We will make our dataset freely available to encourage exploration of more expressive models. 




\section{Conclusions and Future Work}

Answer passage retrieval within topically relevant documents shows unique challenges not present in other QA collections. Until this collection, there were no previous answer passage retrieval collections available that were suitable for the exploration of deep learning models. We presented this new collection and benchmarks to provide an openly available resource so that others can extend our research on non-factoid answer passage retrieval. For the future work, we will study more different neural architectures for non-factoid answer passage retrieval. Answer summarization for non-factoid QA is also an interesting direction to explore.

\section{Acknowledgements}
This work was supported in part by the Center for Intelligent Information Retrieval, in part by NSF \#IIS-1160894, and in part by NSF grant \#IIS-1419693. Any opinions, findings and conclusions or recommendations expressed in this material are those of the authors and do not necessarily reflect those of the sponsor.


\enlargethispage{-2\baselineskip}

\bibliographystyle{ACM-Reference-Format}
 \bibliography{reference}  

\end{document}